# Bragg diffraction of higher orders on oblique helicoidal liquid crystal structures

H. Bogatyryova[a], V. Chornous[b], L. Lisetski[c], I. Gvozdovskyy[a]*

[a]Department of Optical Quantum Electronics, Institute of Physics of the National Academy of Sciences of Ukraine, Kyiv, Ukraine;

[b]Department of Medical and Pharmaceutical Chemistry, Bukovinian State Medical University, Chernivtsi, Ukraine;

[c]Department of Nanostructured Materials, Institute for Scintillation Materials of STC "Institute for Single Crystals" of the National Academy of Sciences of Ukraine, Kharkiv, Ukraine;

Institute of Physics of the National Academy of Sciences of Ukraine, 46 Nauky ave., Kyiv, 03028, Ukraine, telephone number: +380 44 5250862, *e-mail: igvozd@gmail.com

For the oblique helicoidal structure of the chiral twist-bend nematic-forming mixture of CB7CB/CB6OCB/5CB doped by light-sensitive chiral compound based on azo-fragment, two consequent states of Bragg reflection of the light in the visible spectral range from 400 to 750 nm were experimentally observed in the course of decreasing the applied electric field. These states were assigned to different orders of Bragg's diffraction. While the second and third orders of Bragg's diffraction can be easily observed in conditions of standard electrooptic cell, the first order could be revealed only at very high electric field near the "red" edge of the visible spectral range. For the second and third orders of diffraction there is a narrow range of electric field where both diffraction orders are observed (so-called hybrid state), with transmittance spectra of the $Ch_{OH}$ structure showing two peaks at the short-wavelength and the long-wavelength edges of the visible spectral range. Model calculations of Bragg's reflection of light were carried out using literature values of elasticity constants and dielectric permittivity, and the calculated plots of the selective reflection maximum vs. voltage were in good agreement with experimental data.

1. **Introduction**

The phenomenon of selective Bragg reflection of light (BRL) in the visible spectral range is one of the most important advantages of liquid crystals with helicoidal structure (*e.g.* chiral nematics, blue phases and oblique helicoidal structure of the chiral twist-bend nematics), which can be used as promising materials for various applications.

The adding of a certain amount of chiral dopant (ChD) to the nematic host leads to the inducing of helicoidal structure (Figure 1a) characterized by the helical pitch length $P$ that can change within a wide range from nanometres to centimetres. [1,2]

In the case of selective BRL in the visible spectral range, on the one hand, the diffracted light has the wavelength $\lambda$, which is related to the period $\Lambda$ of the diffraction grating as follows: [1,3]

$$m_i \lambda = 2\Lambda \times \sin(\alpha), \quad (1)$$

where $m_i$ = 1, 2, 3,…k and $\alpha$ are the diffraction order and incidence angle, respectively.

On the other hand the maximum of the reflected wavelength $\lambda_{max}$ and the helical pitch $P_0$ are interrelated as follows: [1,3]

$$\lambda_{max} = <n> \times P_0, \quad (2)$$

where $<n> = (n_o + n_e)/2$ is the average refractive index, determined by the ordinary $n_o$ and extraordinary $n_e$ refractive indices of the medium. [1]

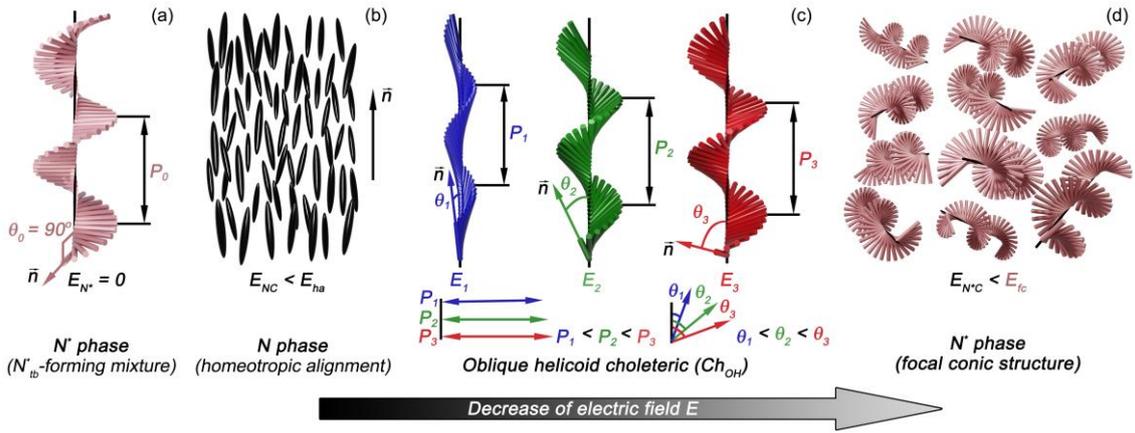

Figure 1. Schematic illustration of the electric field-induced changes in the structure of the induced cholesteric (N*) phase of a N*tb-forming mixture: (a) In the absence of electric field $E$, the local orientation of LC molecules (director $\vec{n}$) is rotating perpendicular to the helical axis with the helical pitch length $P_0$. (b) Homeotropic orientation of director $\vec{n}$ along the applied field when electric field $E_{ha}$ is above the threshold $E_{NC}$ of the nematic-oblique helicoidal cholesteric transition. (c) Ch$_{OH}$ structure with an oblique angle $\theta$ of director $\vec{n}$ under electric field $E < E_{NC}$. The length of the helical pitch $P$ and oblique angle $\theta$ increase with decreasing applied electric field $E$. (d) Ch$_{OH}$ - N* focal conic transition, when certain threshold electric field $E_{fc} < E_{N*C}$ is achieved.

Recently, great attention has been paid to twist-bend nematics (N$_{tb}$), characterized by low value of the bend elastic constant $K_{33}$ in comparison to the twist elastic constant $K_{22}$, as predicted by R. Meyer. [4] The molecules of N$_{tb}$ phase-forming compounds should have banana-like shapes, as was analytically demonstrated by I. Dozov [5] and S. Shamid et al. [6] According to these studies, [4-6] in the N$_{tb}$ phase the director $\vec{n} = (\sin\theta \times \cos\varphi, \sin\theta \times \sin\varphi, \cos\theta)$, i.e., is tilted to the helical axis Z at an angle of $0 < \theta < \pi/2$ and twisted at an azimuthal angle $\varphi$ in the XY-coordinate plane, thus forming an oblique helicoidal structure characterized by the pitch $P$ of the helix.

Adding a small quantity of a chiral dopant (ChD) to the N$_{tb}$ phase leads to formation of the chiral twist-bend nematic (N*$_{tb}$), which was studied in detail. [7-9] For these mixtures at temperature above N*$_{tb}$ phase the transition to the N* phase occurs (Figure 1a). Upon application of the alternating electric field $E$ along the helical axis, the N phase (Figure 1b) is formed above a certain threshold $E_{NC}$ which with the decrease of electric field transforms to the peculiar form of N* phase, which is known as the oblique helicoidal structure (denoted by the abbreviation Ch$_{OH}$ for short) (Figure 1c). [8] Under certain conditions, *e.g.* concentration *C* of ChD, temperature *T*, light or applied electrical field *E* for both the helicoidal structure of cholesteric [10-14] and Ch$_{OH}$ structure of the N$^*_{tb}$-forming mixture [8,9,15-18] selective BRL can be observed in the visible spectral range.

There exists a certain threshold electric field $E_{NC}$ when the oblique helicoidal structure is transformed to the homeotropically aligned nematics (Figure 1a). This threshold field can be expressed as follows: [4,8,9]

$$E_{NC} = \frac{2\pi \cdot K_{22}}{P_0 \cdot \sqrt{\varepsilon_0 \cdot \Delta\varepsilon \cdot K_{33}}}, \quad (3)$$

where $\varepsilon_0$ is the constant of vacuum permittivity and $\Delta\varepsilon$ is the dielectric anisotropy of the N$^*_{tb}$-forming mixture, $K_{22}$ and $K_{33}$ are twist and bend elastic constants, respectively; $P_0$ is the length of cholesteric pitch of N$^*$ phase of the N$^*_{tb}$-forming mixture in the absence of applied electric field.

The Ch$_{OH}$ structure shows selective BRL in the visible spectral range under certain conditions of electric field *E* and temperature *T*. The wavelength of the BRL maximum $\lambda$ can be tuned by changing the applied field *E*, [8,9,15,16] frequency of electric field *f* [16] and temperature *T*. [17,18] This makes them attractive materials for different applications in optics and LC display manufacturing.

When the electric field *E* is gradually decreased, a new threshold value of electric field ($E_{N^*C}$) can be observed, at which the Ch$_{OH}$ structure transforms to conventional helicoidal structure (*i.e.* $\theta = 90°$) with focal conic texture (Figure 1c), which is accompanied by intense scattering of light. This threshold electric field can be expressed as: [4,8]

$$E_{N^*C} \approx E_{NC} \cdot \frac{K_{33}}{K_{22} + K_{33}} [2 + \sqrt{2 \cdot (1 - \frac{K_{33}}{K_{22}})}] \quad (4)$$

In the case of the N$^*_{tb}$-forming phase a small amount of ChD is used to obtain the BRL in the visible spectral range under alternating electric field *E*. The adding of a light-sensitive achiral mesogen [19,20] or chiral azo-compounds [21,22] subject to *E-Z* photoisomerization of molecules to the $N^*_{tb}$ phase makes it possible to obtain the light-tuning of the wavelength of selective BRL for $Ch_{OH}$ structure. In this process, both the tilt angle $\theta$ of director $\vec{n}$ and the pitch $P$ are changed.

In this work we will show two sequential states of BRL in the wide visible spectral range, with two separate $\lambda(E)$ plots registered at high and lower voltages applied to $N^*$ phase of $N^*_{tb}$-forming mixture. We will try to understand the reason of the appearance of these two experimentally observed states of light reflection in the visible spectral range.

## 2.     Materials and methods

As a $N^*_{tb}$-forming mixture with selective BRL in the visible range of spectrum at room temperature, we used the four-component mixture recently studied in detail. [21] The basic formulation is the three-component mixture of nematics, namely two achiral nematic dimers 1',7''-bis-4-(4-cyanobiphen-4'-yl)heptane (CB7CB) and 1-(4-cyanobiphenyl-4'-yloxy)-6-(4-cyanobiphenyl-4'-yl)hexane (CB6OCB), as well as nematic 4-pentyl-4'-cyanobiphenyl (5CB), added to lower the temperatures of the mesophase. The twist-bend nematic CB7CB was synthesized in Bukovinian State Medical University (Chernivtsi, Ukraine). It forms uniaxial nematic N phase in the temperature range of 103 - 116 ºC, which is in agreement with data presented in Ref. [23-25] The twist-bend nematic CB6OCB was obtained from Synthon Chemicals GmbH & Co (Wolfen, Germany), forming uniaxial N phase in the temperature range of 109 - 157 ºC, [26-28] while the nematic 5CB was synthesized and purified before using at the STC "Institute of Single Crystals" (Kharkiv, Ukraine). The weight ratio of CB7CB:CB6OCB:5CB in the $N_{tb}$-forming mixture was 39:19:42, respectively. The temperatures of the phase transitions of this mixture were measured in Ref. [21] both on cooling ($N_{tb}$ 38.6°C N 78.2 °C Iso) and on heating (Iso 77.2 °C N 37.2°C $N_{tb}$).

The temperature dependences of the elastic and dielectric constants, as well as refractive indices of CB7CB were carefully measured in Ref., [29] which we used for our approximate calculations.

To induce the $N^*_{tb}$ phase for which the $Ch_{OH}$ structure arises electric field $E$ at room temperature, we used the light-sensitive chiral (1$R$,2$S$,5$R$)-2-isopropyl-5-methylcyclohexyl-4-{($E$)-[4-(hexanoyloxy)phenyl]diazenyl}benzoate (ChD-3816 for short), [21] synthesized in Bukovinian State Medical University (Chernivtsi, Ukraine).

To study the features of the selective BRL under applying of the electric field $E$ and UV irradiation, taking into account the studies carried out in Ref,[21] the three-component $N_{tb}$-forming mixture was doped by ChD-3816 with concentration 10 wt. %.

The planar alignment of $N^*_{tb}$ mixture is provided by means of the polyimide PI2555 (HD MicroSystems, USA), [30] which was unidirectionally rubbed $N_{rubb} = 15$ times to ensure strong azimuthal anchoring. [31]

PI2555 was dissolved in the *n*-methyl-2-pyrrolidone in proportion 10:1, and further, to obtain a thin polyimide film, this solution was spin-coated (6800 rpm, 20 s) on glass substrates covered with indium tin oxide (ITO) layer.

LC cells were assembled with two substrates rubbed in opposite directions. The thickness $d$ of LC cells were set by 20 µm in diameter Mylar spacers and measured by means of the transmission spectrum of the empty cell using an Ocean Optics USB4000 spectrometer (Ocean Insight, USA, California). The LC cells were filled by capillary action at 58 °C, *i.e.*, above the temperature of the isotropic phase (Iso) transition $T_{Iso} = 56.9$ °C, [21] and then slowly cooled.

Transmission spectra of the $Ch_{OH}$ structure, characterized by the selective BRL in the visible spectral range, were recorded by means of the spectrometer Ocean Optics USB4000 in the range of wavelengths 400 – 750 nm.

To change the wavelength of BRL in the visible spectral range, we applied the alternating voltage within the range 0 - 92 V and frequency 1 kHz to the LC cell filled by the $N^*_{tb}$-forming mixture.

To stabilize the temperature of the $N^*_{tb}$-forming mixture containing 10 wt. % of ChD-3816, we used a thermostable heater based on a temperature regulator MikRa 603 (LLD 'MikRa', Kyiv, Ukraine) equipped with a platinum resistance thermometer Pt1000 (PJSC 'TERA', Chernihiv, Ukraine). The temperature measurement accuracy was ± 0.1 °C/min. The studies of the selective BRL were carried out at constant temperature 27 °C.

## 3. Results and discussions

*3.1. Experimental observation of two states of the selective Bragg reflection from $Ch_{OH}$ structure*

In this section we will describe our experimental results on the influence of the applied electrical field $E$ (or voltage $U = E \times d$) upon the wavelength of the selective BRL of the $Ch_{OH}$ structure.

The dependence of wavelength $\lambda$ of the selective BRL on electric field $E$ applied to the $Ch_{OH}$ structure is typical, as described in detail elsewhere. [8,9] The experimentally observed red-shift of $\lambda$ caused by the decreasing of the applied electric field $E$ (Figure 2), reflecting the corresponding increase in the helical pitch values. The change of transmittance spectrum of the selective BRL is shown in Figure 3. As distinct from, [8,9,15-17,19,20,22] we experimentally obtained at least two sequential states of changes in the reflected wavelength $\lambda$ in the wide spectral range caused by the decrease of electric field $E$ of the $Ch_{OH}$ structure.

At the threshold electric field $E_{NC}$ (Equation (3)) the homeotropically aligned N phase transforms to the $Ch_{OH}$ structure. It has been experimentally found that there is a state in a certain range of "high" electric field $E$ ($E_2^{\lambda_2} \leq E_{high} \leq E_1^{\lambda_1}$) when we observe selective BRL in the visible spectral range 420 - 720 nm (*i.e.* $\lambda_1$ = 420 nm and $\lambda_2$ = 720 nm). This state we will denote as the State 1 (solid red spheres, curve 1, Figure 2).

According to the Equation (3), the State 1 is characterized by the typical red-shift of the selective BRL wavelength with the decrease of the electric field $E$. The Figure 3a shows the transformation of the transmission spectrum of the $Ch_{OH}$ structure under electric field. According to, [8,9] the reason of the red-shift of wavelength is the increase of both the tilt angle $\theta$ of director $\vec{n}$ of the $Ch_{OH}$ structure and the pitch $P$ of helix with the decrease of voltage $U$. For our experimental conditions, *i.e.* when the light incidence angle $\alpha$ was $\pi/3$ and the thickness $d$ of LC cell was 25.2 μm, the selective BRL in the visible spectral range is observed when applied electric field $E$ varies from 2.32 to 1.59 V/μm (*i.e.* $U$ ~ 59 V - 40 V).

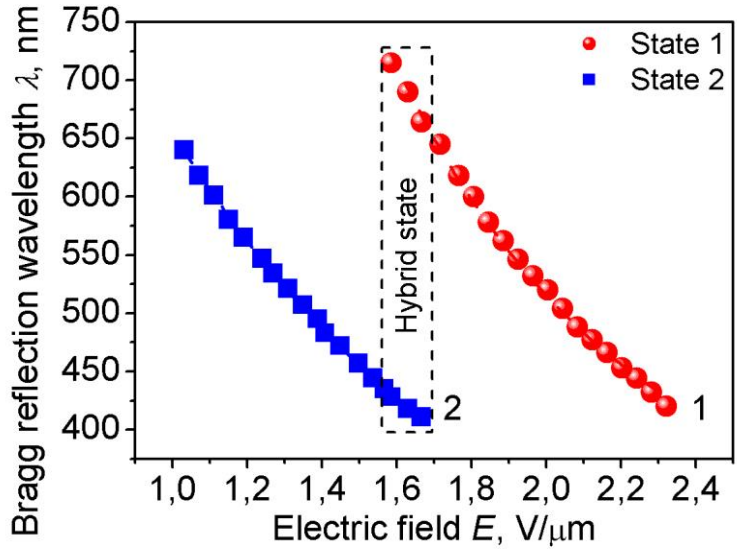

Figure 2. Wavelength $\lambda$ of the peak of selective BRL as function of applied electric field $E$. The sequential stages of the peak red-shift are: (1) State 1 at high electric field $E_{high}$ (solid red spheres, curve 1) and (2) State 2 at low electric field $E_{low}$ (solid blue squares, curve 2). The hybrid state region is marked by dashed black rectangle. The incidence angle $\alpha$ of light was $\pi/3$. The frequency $f$ of applied electric field was 1 kHz. The temperature $T$ of $Ch_{OH}$ structure was 27 °C.

On the subsequent further decreasing of electric field $E$ we additionally found another state, which was also characterized by the selective BRL in the visible spectral range (solid blue squares, curve 2, Figure 2). Similarly to the $Ch_{OH}$ structure in high electric field $E_{high}$, the selective BRL in the visible spectral range 410 - 650 nm (*i.e.* $\lambda_1 =$ 410 nm and $\lambda_2 = 650$ nm) is also observed at low electric field $E$ ( $E_2^{\lambda_2} \leq E_{low} \leq E_1^{\lambda_1}$ ). In this case the red-shift of wavelength with the decrease of electric field $E_{low}$ within the range from 1.67 to 1.03 V/μm (*i.e.* $U \sim 42 - 26$ V) is also observed (Figure 3b). This state of $Ch_{OH}$ structure we will denote as the State 2.

In the case of State 2 there is a threshold electric field $E_{N*C} \leq E_{fc}$ (Equation (4)), when the $Ch_{OH}$ structure transforms to conventional helicoidal structure of the $N^*$ phase (director $\vec{n}$ is perpendicularly oriented to the helicoid axis, *i.e.* $\theta = \pi/2$), characterized by focal conic structure (Figure 1d). For our experimental conditions, at electric field $E_{fc}$ < 1.03 V/μm (*i.e.* $U < 25.9$ V) the intense scattering of light is observed.

It is important to note that the decrease of electric field applied to the $Ch_{OH}$ being in the State 1 leads to the appearance of the hybrid state, when the State 1 and State 2 are observed together. The hybrid state there exists within narrow electrical field range from 1.59 to 1.67 V/μm (*i.e. U* ~ 40 - 42 V), marked by the dashed black rectangle as shown in the Figure 2. Because hybrid state is not "long-lived", then with decreasing of electric field *E*, the switching between long-wavelength State 1 to the short-wavelength State 2 takes place.

The Figure 3 shows sequential red-shift of the transmittance spectra (marked by numbers from 1 to 10) with the decrease of electric field *E* for both State 1 and State 2. It should be noted that for the $Ch_{OH}$ structure the inverse blue-shift of the peak of transmittance spectrum with increasing electric field *E* within range from $E_{low}$ to $E_{high}$ is also observed. However, it was found that the storage of $Ch_{OH}$ structure during few minutes at low electric field $E_{low}$ (*i.e.* in the State 2 characterized by large pitch $P_3$ of helix of the oblique helicoid and angle *θ* of director $\vec{n}$ as schematically shown in Figure 1c) can lead to difficulties in further observation of the red-shift of wavelength and the selective BRL (so-called metastable state) in general. It is accompanied by transition from $Ch_{OH}$ to $N^*$ phase with focal conic structure (Figure 1d).

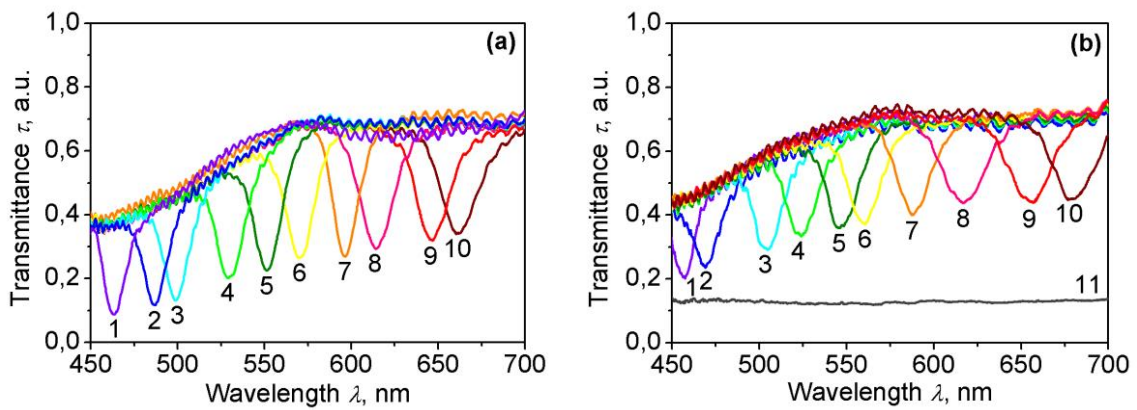

Figure 3. The change of transmittance spectrum of the $Ch_{OH}$ structure in the State 1 (a) and State 2 (b) with the decrease of applied electric field *E*. The incidence angle *α* of light was π/3. The frequency *f* of applied electric field was 1 kHz. The temperature *T* of $Ch_{OH}$ structure was 27 °C.

The Figure 3b also shows the transmittance of $N^*$ phase with focal conic structure (spectrum 11), the appearance of which is characterized by intense scattering of light. Due to the occurrence of light scattering no selective BRL is observed after subsequent increase of the electric field. To observe the selective BRL in the visible spectral range, the applied electric field *E* again should be increased to the value of threshold $E_{NC}$ (Equation (3)), and further it should be slowly decreased.

The Figure 4 shows experimentally obtained transmission spectra of the $Ch_{OH}$ structure in the hybrid state, characterized by the presence of two peaks in contrary to spectra presented in the Figure 3. This is made possible by the recording wider spectral range from 400 to 750 nm when changing the electrical field within narrow range (*i.e.* from 1.59 to 1.67 V/μm), which provides the observation at the same time of the State 1 and State 2.

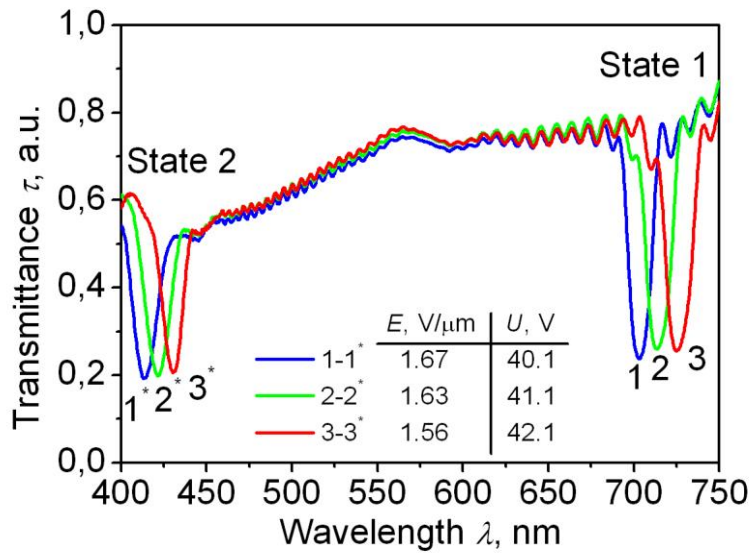

Figure 4. The electrical field induced transformation of transmittance spectrum of the $Ch_{OH}$ structure in hybrid state, when the State 1 and State 2 appear at the same time. With decreasing electric field, the pairs of wavelengths of the BRL in the visible range for the each transmittance spectrum are: (1-1*) 700 –415 nm, (2-2*) 715 - 420 nm and (3-3*) 725 - 430 nm. The incidence angle *α* of light and thickness of LC cell were $\pi/3$ and 25.2 μm, respectively. The frequency *f* of the applied electric field was 1 kHz. The temperature *T* of $Ch_{OH}$ structure was 27 °C.

As can be seen from Figure 4, the dynamic of transmission spectrum 1-1* is characterized by of sequential red-shifts of wavelengths of the BRL (spectra 2-2* and 3-

3[*]) for both the long-wavelength (*i.e.* from 700 nm to 725 nm) and the short-wavelength (*i.e.* from 415 nm to 425 nm) spectral ranges.

The Figure 5a shows the change of full width at half minimum $\Delta\lambda_{FWHM}$ (FWHM for short) of the transmittance spectrum of Ch$_{OH}$ structure as the function of applied electric field *E*. For both states the monotonous decrease of $\Delta\lambda_{FWHM}$ with the increase of electric field *E* is observed. For both the State 1 (solid red squares, curve 1) and the State 2 (solid blue circles, curve 2) when electric field is applied, the function of $\Delta\lambda_{FWHM}(E_{high})$ qualitatively agrees with results obtained elsewhere. [15,17] The reflection intensity $100-\tau_{min}$ at the maximum wavelength of selective BRL as function of applied field is shown in Figure 5b for both states of Ch$_{OH}$ structure. It should be noted that the intensity of BRL in State 2 is substantially lower than in State 1 at the same peak wavelengths.

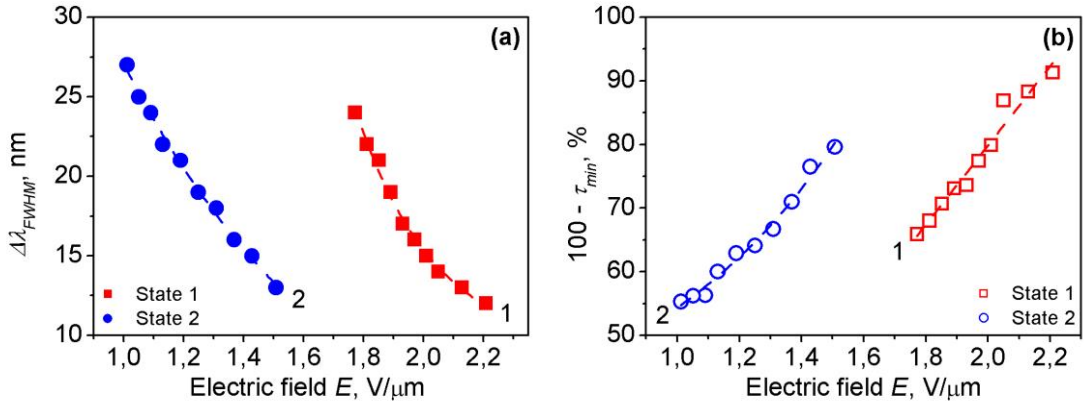

Figure 5. Full width at half minimum $\Delta\lambda_{FWHM}$ of the transmittance spectrum of Ch$_{OH}$ structure in the State 1 (solid red squares, curve 1) and State 2 (solid blue circles, curve 2) (a), and reflection intensity at the maximum wavelength of the selective BRL (b) as functions of applied electric field *E*. The incidence angle *α* of light was $\pi/3$. The frequency *f* of the applied electric field was 1 kHz. The temperature *T* of Ch$_{OH}$ structure was 27 °C.

In addition, the width of the bandwidth area is approximately the same for both states. As a result the Ch$_{OH}$ structure displays the same colours, each observed at two different values of electric field *E*, as can be seen in Figure 6.

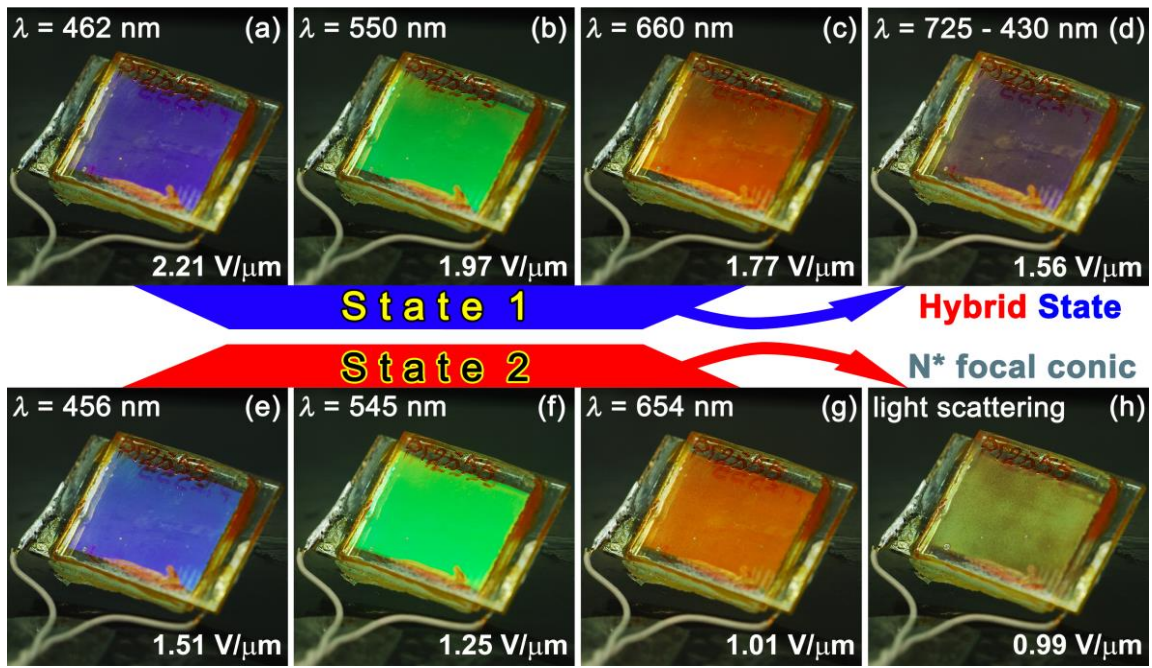

Figure 6. Electrical tuning of the wavelength of Bragg reflection of light in the visible spectral range for the State 1 at: (a) - 2.21 V/µm ($\lambda$ = 462 nm), (b) - 1.97 V/µm ($\lambda$ = 550 nm) and (c) - 1.77 V/µm ($\lambda$ = 660 nm), the hybrid state at: (d) – 1.56 V/µm ($\lambda$ = 750 – 430 nm), the State 2 at: (e) - 1.51 V/µm ($\lambda$ = 456 nm), (f) - 1.25 V/µm ($\lambda$ = 545 nm) and (g) - 1.01 V/µm ($\lambda$ = 654 nm) and $N^*$ phase with focal conic structure at: (h) - 0.99 V/µm (light scattering). The incidence angle $\alpha$ of light was $\pi/3$. Thickness of LC cell was 25.2 µm. The frequency $f$ of the applied electric field was 1 kHz. The temperature $T$ of $Ch_{OH}$ structure was 27 °C.

However, it is also important to note that for the same colours (*i.e.* reflected wavelengths), when $Ch_{OH}$ structure being in the State 2 (opened blue circles, curve 2), the reflection light is less intense than it is for the State 1 (opened red squares, curve 1) as can be seen from Figure 4b.

*3.2. What is the origin of different states observed in the voltage-dependent selective Bragg reflection of light of the $Ch_{OH}$ structure?*

In this section we will try to find out the origin of the experimentally observed two states of selective BRL of $Ch_{OH}$ structure under electric field in the visible spectral range. Our consideration will be based on calculations of the change of wavelength of the BRL taking into account Equations (1) - (4).

According to the Bragg's law the maximum of reflected wavelength for various orders of diffraction is expressed by Equations (1).

In the particular case when $\alpha = \pi/2$, the period of grating $\Lambda$ and the length of helix pitch $P_0$, accounting for the Equation (2), are interconnected as follows:

$$\begin{cases} \Lambda_{m=1} = <n> \times \dfrac{P_0}{2} \\ \Lambda_{m=2} = <n> \times P_0 \\ \Lambda_{m=3} = <n> \times \dfrac{3}{2} P_0 \end{cases} \quad (5)$$

where $m_i = 1$, 2 and 3 are the respective orders of diffraction. We limit ourselves to just three orders of diffraction in our considerations.

For the general case, when $0 < \alpha \leq \pi/2$, it follows from the Equation (2) and Equation (5) that the maximum of wavelength of the selective BRL and the threshold electric field $E_{NC}$ are interconnected as:

$$\lambda = \frac{1}{m_i} \sin(\theta) \cdot \frac{2\pi \cdot K_{22} \cdot <n>}{E\sqrt{\varepsilon_0 \cdot \Delta\varepsilon \cdot K_{33}}} \quad (6)$$

The Figure 7 shows curves of calculated function $\lambda(E)$ for three orders of diffraction (Equation (6)) by taking account of some experimental data (*e.g.* thickness *d* of the LC cell, incidence angle $\alpha$ of light and applied voltage *U*) and parameters of liquid-crystalline medium. The calculations are carried out for the general case (*i.e.* the wide spectral range from ultraviolet to infrared) and the partial case (*i.e.* in the visible spectral range from 400 to 750 nm).

For example, as can be seen in the Figure 7a, at a certain value of electric field *E* various orders of Bragg diffraction can be observed simultaneously. But here we will limit ourselves only to the visible spectral range from 400 to 750 nm (*i.e.* the range selected by the dashed green rectangle as shown in the Figure 7a) used in our experiments. This selected range in scale-up is shown in the Figure 7b.

We can see that at all electric filed *E* values not all the orders of diffraction (*i.e.* the selective BRL in the visible range) can be observed simultaneously. The first diffraction order (solid black curve 1) can be observed at very high electric field (*i.e.* at large value of applied voltage $U$), while the second (dashed blue curve 2) and third (dash-doted red curve3) orders of Bragg diffraction can be obtained at smaller values of voltage $U$ (*i.e.* at low fields $E = U/d$).

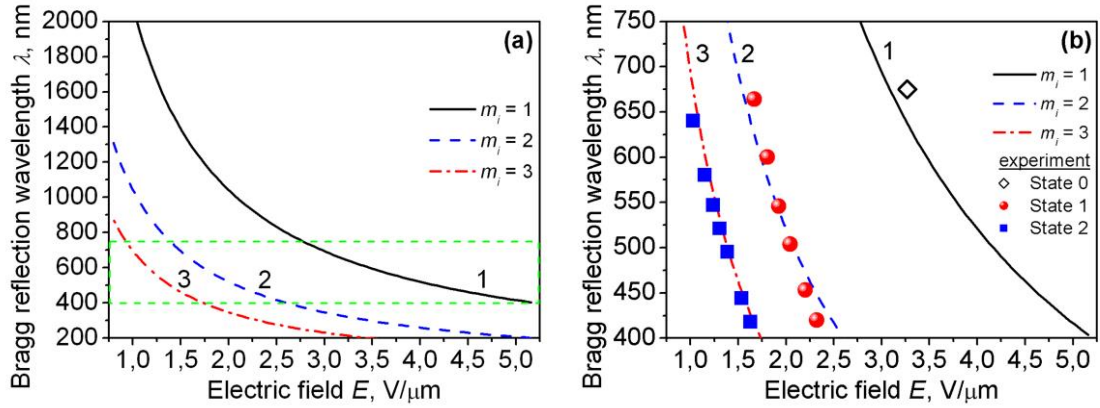

Figure 7. The electrical field dependence of the maximum wavelength of the selective Bragg reflection calculated for three orders of diffraction in the case of: (a) wide (*i.e.* 200 – 2000 nm) and (b) visible (*i.e.* 400 – 750 nm) spectral ranges. The order of diffraction is: 1) $m_i = 1$ (solid black curves 1), 2) $m_i = 2$ (dashed red curves 2) and 3) $m_i = 3$ (dash-doted red curves 3). The calculation was carried out for the incidence angle $\alpha = \pi/3$ and 25.2 µm LC cell. Experimental points are plotted on each curve for the: 1) State 0 (opened black diamond), 2) State 1 (red spheres) and 3) State 2 (solid blue squares). In case of the State 0 the incidence angle $\alpha = 65°$ was used, while for the State 1 and State 2 $\alpha = 60°$.

By comparing the experimentally obtained two states of selective BRL (Figure 2) with calculated function $\lambda(E)$ for the partial case (Figure 7b) we can conclude that electrical tuning of wavelength of the BRL for the certain experimental conditions (*e.g.* thickness $d$ of LC cell, incidence angle $\alpha$ of light, composition of mixture having certain value of elastic constant $K_{22}$, $K_{33}$ and average refractive index $<n>$) can be characterized by the observation of different orders of the Bragg diffraction.

The curve 1 of Figure 7b shows the first order of Bragg diffraction of the $Ch_{OH}$ structure, which can be obtained at very high electric field within range from 2.78 to 5.16 V/µm (*i.e.* at $U \sim 70.1 – 130$ V). According to the Equation (5), the diffraction of first order will also be observed, when the periodicity of $Ch_{OH}$ structure corresponds to the following condition: $\Lambda_{m=1} = <n> \cdot P_0/2$. Let us denote the state of Ch$_{OH}$ structure with diffraction of first order as the State 0.

Analyzing the calculated data shown in the Figure 7b, we can assume that the diffraction of the first order could also be experimentally observed. Accounting for the real technical ability of the used generator to change the applied voltage $U$ in the range from 0 to 92 V with frequency $f$ = 1 kHz, we can conclude that the Bragg's diffraction of first order should be observed within limited range of spectrum (*i.e.* 570 - 750 nm at electric field changes from 3.65 - 2.78 V/µm). It should also be noted that application of very high electric fields may lead to electrical breakdown of the LC cell.

However, here our main goal was to experimentally detect the Bragg's diffraction of first order (*i.e.* when Ch$_{OH}$ structure would be in the State 0), even at very high electrical field. By altering the incidence angle $\alpha = \pi/2.8$, the Bragg's diffraction of first order at $E_{high}$ within 2.98 - 3.65 V/µm (*i.e.* 75.1 - 92 V) was obtained.

The Figure 8 shows the recorded transmission spectrum of our Ch$_{OH}$ structure in the State 0 at very high electrical field $E_{high}$ about 3.27 V/µm (or 82.5 V). In order to facilitate observation of the Bragg's diffraction of first order, the incidence angle was changed to $\alpha \sim \pi/2.8$, as distinct from conditions in the Figure 3. This experimental point of Bragg's diffraction of the first order is also shown in the Figure 7b by the black sphere symbol. For these experimental conditions the maximum wavelength $\lambda_{max}$ of Ch$_{OH}$ structure in the State 0 was about 675 nm. In this case the selective BRL is observed, on the one hand, near the long-wavelength edge of the visible spectral range from 670 - 750 nm and, on the other hand, at very high electric field, which can be considered as a certain hindrance to actually using the State 0.

The Figure 7b shows that decrease of electric field $E$ within ranges 1.39 - 2.58 V/µm (*i.e.* $U \sim$ 35 – 65 V) and 0.93 - 1.75 V/µm (*i.e.* $U \sim$ 23 – 44 V) allows us to observe the diffraction of the second and third orders. By comparing the experimental obtained states (Figure 2) and calculated (Figure 7) we could conclude that the above-noted State 1 and State 2 correspond to the second and third order of Bragg's diffraction, respectively. It should be noted that the Figure 7b also shows some experimental points moved from Figure 2 and their satisfactory agreement with calculated curves is observed.

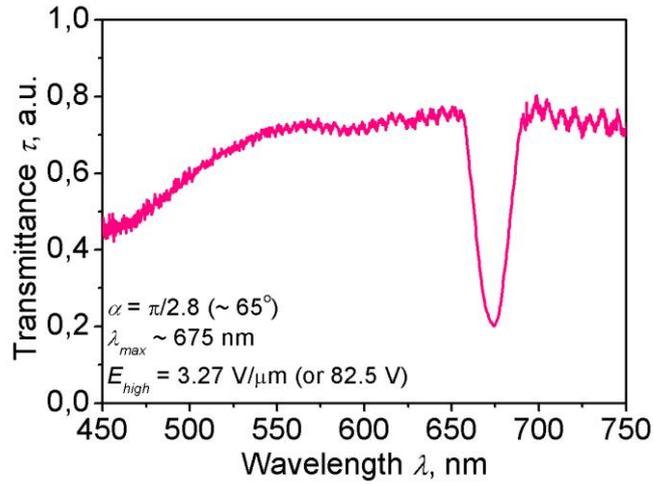

Figure 8. The transmission spectrum of the studied Ch$_{OH}$ structure under very high electric field $E_{high}$ ~ 3.27 V/μm applied to 25.2 μm LC cell. Incidence angle $α$ of light was about 65°. The minimum of transmittance spectrum is at wavelength about 675 nm.

According to Equation 5, the State 1 (*i.e.*, diffraction of the second order) and the State 2 (*i.e.*, diffraction of the third order) are possible, when respective conditions of $Λ_{m=2} = <n>·P_0$ and $Λ_{m=3} = <n>·3P_0/2$ are realized.

From the Figure 7b it follows that if the wavelength range is within 400 - 750 nm, then the first order of diffraction is observed on its own and does not overlap with other orders of Bragg's diffraction. However, for the diffraction of second and third orders the overlapping within narrow electric field range from 1.5 to 1.65 V/μm (or $U$ ~ 38 - 42 V) appears. When the electric field $E$ is sufficiently decreased, two peaks appear together - at long-wavelength range for the second order of diffraction and short-wavelength spectral range for the third order of diffraction. The experimental data obtained for the hybrid state (Figure 2) displaying two peaks of transmittance spectrum of the Ch$_{OH}$ structure (Figure 4) is in good agreement with calculation shown in the Figure 7. In addition the switch-over of the order of the Bragg's diffraction with the decrease in the electric field E is accompanied with the repeated sequence of colour changes *i.e.*, variation of the pitch $P$ of Ch$_{OH}$ helix) through hybrid state, as it was observed experimentally for the State 1 and State 2 (Figure 6).

As was mentioned above, at very low electric field $E_{N*C}$ (Equation (4)) the transition of Ch$_{OH}$ structure to the N$^*$ phase with focal conic texture occurs (Figure 1d). The Figure 9 shows the threshold electrical field $E_{N*C}$ depending on the threshold field

$E_{NC}$ (Equation (3)) which allows to determine the range of electrical field $E$ where the $Ch_{OH}$ structure of $N^*_{tb}$-forming mixture displays the selective BRL in the wide visible spectral range. Since the $N^*_{tb}$-forming mixture characterized by the certain set of elastic constants (*i.e.* $K_{22}$ and $K_{33}$), different refractive indices (*i.e.* $n_o$ and $n_e$), dielectric anisotropy $\Delta\varepsilon$, concentrations $C$ of the ChD used *etc.*, the linear function $E_{N*C}(E_{NC})$ will be individual for each particular case.

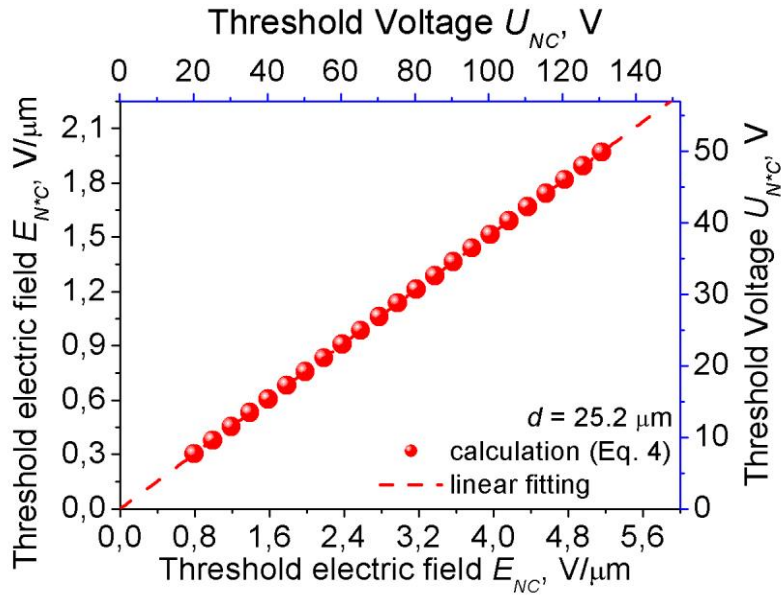

Figure 9. The linear dependence of the threshold electric field $E_{N*C}$ (or voltage $U_{N*C}$, when the $N^*_{tb}$ – $N^*$ phase transition occurs) on the threshold electric field $E_{NC}$ (or voltage $U_{NC}$, when the $N$ – $N^*$ phase transition of $N^*_{tb}$-forming mixture takes place).

As recently found in, [17] the $Ch_{OH}$ structure can exist, when inequality $E_{N*C}/E_{NC} < 1$ is satisfied. In our case this ratio is about 0.38. By knowing the value of threshold of the electric field $E_{NC}$ of the $N$ – $Ch_{OH}$ structure transition, owing to the function $E_{N*C}(E_{NC})$ as in the Figure 9, we can determine the value of threshold of electric field $E_{N*C}$, when $Ch_{OH}$ - $N^*$ phase transition occurs. On the one hand it can be useful for application, namely to predict the amount of observed states for certain electric field range, where $Ch_{OH}$ structure, possessing the selective BRL in the visible spectral range, exists. On the other hand, by taking into account the Figure 7b, for certain value of electric field we can find both the maximum wavelength of the BRL in all range of the existence Ch$_{OH}$ structure and the order of Bragg diffraction, hybrid state possessing by two wavelengths of BRL at the same time (Figure 4).

**Conclusions**

Selective Bragg reflection of the light was studied for Ch$_{OH}$ structure of N$^*_{tb}$-forming mixture CB7CB/CB6OCB/5CB/ChD-3816 under electric field.

We tried to describe and explain the phenomenon of the existence of the two observed states of selective Bragg reflection in the wide visible spectral range from 400 to 750 nm, where the wavelength of maximum reflection could be tuned by decreasing the electric field applied to the Ch$_{OH}$ structure.

We concluded that observed states are related to different orders of Bragg's diffraction on the oblique helicoidal structure, and in this case the reflection of higher orders is manifested much more clearly than with helicoidal structures of conventional cholesterics.

While the second and third orders of Bragg's diffraction can be easily observed in conditions of standard electrooptic cell, the first order of Bragg's diffraction can be revealed only at very high electric field. It was experimentally shown that the first order of diffraction, as predicted by calculations, could be observed near the edge of the visible spectral range.

For the second and third orders of diffraction there is a narrow range of electric field where both diffraction orders are observed (so-called hybrid state). It has been experimentally found that the transmittance spectra of the Ch$_{OH}$ structure in hybrid state shows two peaks at the short-wavelength and the long-wavelength edges of the visible spectral range.

The reported features of Ch$_{OH}$ structure of N$^*_{tb}$-forming mixtures seem to be promising for their use in electrooptic applications.